\documentstyle[12pt]{article}
\begin{document}
\begin{flushright}
Preprint SSU-HEP-99/04\\
Samara State University
\end{flushright}
\vspace{30mm}
\begin{center}
{ {\bf PROTON POLARIZABILITY AND THE LAMB\\
SHIFT IN MUONIC HYDROGEN}\\
R.N.~Faustov\\
Scientific Council "Cybernetics" RAS,\\
117333, Moscow, Vavilov, 40, Russia\\
A.P.~Martynenko\\
Department of Theoretical Physics, Samara State University\\
443011, Samara, Pavlov, 1, Russia}
\end{center}

\newdimen\Lengthunit	   \Lengthunit	= 1.5cm
\newcount\Nhalfperiods	   \Nhalfperiods= 9
\newcount\magnitude	   \magnitude = 1000

\catcode`\*=11
\newdimen\L*   \newdimen\d*   \newdimen\d**
\newdimen\dm*  \newdimen\dd*  \newdimen\dt*
\newdimen\a*   \newdimen\b*   \newdimen\c*
\newdimen\a**  \newdimen\b**
\newdimen\xL*  \newdimen\yL*
\newdimen\rx*  \newdimen\ry*
\newdimen\tmp* \newdimen\linwid*

\newcount\k*   \newcount\l*   \newcount\m*
\newcount\k**  \newcount\l**  \newcount\m**
\newcount\n*   \newcount\dn*  \newcount\r*
\newcount\N*   \newcount\*one \newcount\*two  \*one=1 \*two=2
\newcount\*ths \*ths=1000
\newcount\angle*  \newcount\q*	\newcount\q**
\newcount\angle** \angle**=0
\newcount\sc*	  \sc*=0

\newtoks\cos*  \cos*={1}
\newtoks\sin*  \sin*={0}

\catcode`\[=13

\def\rotate(#1){\advance\angle**#1\angle*=\angle**
\q**=\angle*\ifnum\q**<0\q**=-\q**\fi
\ifnum\q**>360\q*=\angle*\divide\q*360\multiply\q*360\advance\angle*-\q*\fi
\ifnum\angle*<0\advance\angle*360\fi\q**=\angle*\divide\q**90\q**=\q**
\def\sgcos*{+}\def\sgsin*{+}\relax
\ifcase\q**\or
 \def\sgcos*{-}\def\sgsin*{+}\or
 \def\sgcos*{-}\def\sgsin*{-}\or
 \def\sgcos*{+}\def\sgsin*{-}\else\fi
\q*=\q**
\multiply\q*90\advance\angle*-\q*
\ifnum\angle*>45\sc*=1\angle*=-\angle*\advance\angle*90\else\sc*=0\fi
\def[##1,##2]{\ifnum\sc*=0\relax
\edef\cs*{\sgcos*.##1}\edef\sn*{\sgsin*.##2}\ifcase\q**\or
 \edef\cs*{\sgcos*.##2}\edef\sn*{\sgsin*.##1}\or
 \edef\cs*{\sgcos*.##1}\edef\sn*{\sgsin*.##2}\or
 \edef\cs*{\sgcos*.##2}\edef\sn*{\sgsin*.##1}\else\fi\else
\edef\cs*{\sgcos*.##2}\edef\sn*{\sgsin*.##1}\ifcase\q**\or
 \edef\cs*{\sgcos*.##1}\edef\sn*{\sgsin*.##2}\or
 \edef\cs*{\sgcos*.##2}\edef\sn*{\sgsin*.##1}\or
 \edef\cs*{\sgcos*.##1}\edef\sn*{\sgsin*.##2}\else\fi\fi
\cos*={\cs*}\sin*={\sn*}\global\edef\gcos*{\cs*}\global\edef\gsin*{\sn*}}\relax
\ifcase\angle*[9999,0]\or
[999,017]\or[999,034]\or[998,052]\or[997,069]\or[996,087]\or
[994,104]\or[992,121]\or[990,139]\or[987,156]\or[984,173]\or
[981,190]\or[978,207]\or[974,224]\or[970,241]\or[965,258]\or
[961,275]\or[956,292]\or[951,309]\or[945,325]\or[939,342]\or
[933,358]\or[927,374]\or[920,390]\or[913,406]\or[906,422]\or
[898,438]\or[891,453]\or[882,469]\or[874,484]\or[866,499]\or
[857,515]\or[848,529]\or[838,544]\or[829,559]\or[819,573]\or
[809,587]\or[798,601]\or[788,615]\or[777,629]\or[766,642]\or
[754,656]\or[743,669]\or[731,681]\or[719,694]\or[707,707]\or
\else[9999,0]\fi}

\catcode`\[=12

\def\GRAPH(hsize=#1)#2{\hbox to #1\Lengthunit{#2\hss}}

\def\Linewidth#1{\global\linwid*=#1\relax
\global\divide\linwid*10\global\multiply\linwid*\mag
\global\divide\linwid*100\special{em:linewidth \the\linwid*}}

\Linewidth{.4pt}
\def\sm*{\special{em:moveto}}
\def\sl*{\special{em:lineto}}
\let\moveto=\sm*
\let\lineto=\sl*
\newbox\spm*   \newbox\spl*
\setbox\spm*\hbox{\sm*}
\setbox\spl*\hbox{\sl*}

\def\mov#1(#2,#3)#4{\rlap{\L*=#1\Lengthunit
\xL*=#2\L* \yL*=#3\L*
\xL*=\xscale\xL* \yL*=\yscale\yL*
\rx* \the\cos*\xL* \tmp* \the\sin*\yL* \advance\rx*-\tmp*
\ry* \the\cos*\yL* \tmp* \the\sin*\xL* \advance\ry*\tmp*
\kern\rx*\raise\ry*\hbox{#4}}}

\def\rmov*(#1,#2)#3{\rlap{\xL*=#1\yL*=#2\relax
\rx* \the\cos*\xL* \tmp* \the\sin*\yL* \advance\rx*-\tmp*
\ry* \the\cos*\yL* \tmp* \the\sin*\xL* \advance\ry*\tmp*
\kern\rx*\raise\ry*\hbox{#3}}}

\def\lin#1(#2,#3){\rlap{\sm*\mov#1(#2,#3){\sl*}}}

\def\arr*(#1,#2,#3){\rmov*(#1\dd*,#1\dt*){\sm*
\rmov*(#2\dd*,#2\dt*){\rmov*(#3\dt*,-#3\dd*){\sl*}}\sm*
\rmov*(#2\dd*,#2\dt*){\rmov*(-#3\dt*,#3\dd*){\sl*}}}}

\def\arrow#1(#2,#3){\rlap{\lin#1(#2,#3)\mov#1(#2,#3){\relax
\d**=-.012\Lengthunit\dd*=#2\d**\dt*=#3\d**
\arr*(1,10,4)\arr*(3,8,4)\arr*(4.8,4.2,3)}}}

\def\arrlin#1(#2,#3){\rlap{\L*=#1\Lengthunit\L*=.5\L*
\lin#1(#2,#3)\rmov*(#2\L*,#3\L*){\arrow.1(#2,#3)}}}

\def\dasharrow#1(#2,#3){\rlap{{\Lengthunit=0.9\Lengthunit
\dashlin#1(#2,#3)\mov#1(#2,#3){\sm*}}\mov#1(#2,#3){\sl*
\d**=-.012\Lengthunit\dd*=#2\d**\dt*=#3\d**
\arr*(1,10,4)\arr*(3,8,4)\arr*(4.8,4.2,3)}}}

\def\clap#1{\hbox to 0pt{\hss #1\hss}}

\def\ind(#1,#2)#3{\rlap{\L*=.1\Lengthunit
\xL*=#1\L* \yL*=#2\L*
\rx* \the\cos*\xL* \tmp* \the\sin*\yL* \advance\rx*-\tmp*
\ry* \the\cos*\yL* \tmp* \the\sin*\xL* \advance\ry*\tmp*
\kern\rx*\raise\ry*\hbox{\lower2pt\clap{$#3$}}}}

\def\sh*(#1,#2)#3{\rlap{\dm*=\the\n*\d**
\xL*=\xscale\dm* \yL*=\yscale\dm* \xL*=#1\xL* \yL*=#2\yL*
\rx* \the\cos*\xL* \tmp* \the\sin*\yL* \advance\rx*-\tmp*
\ry* \the\cos*\yL* \tmp* \the\sin*\xL* \advance\ry*\tmp*
\kern\rx*\raise\ry*\hbox{#3}}}

\def\calcnum*#1(#2,#3){\a*=1000sp\b*=1000sp\a*=#2\a*\b*=#3\b*
\ifdim\a*<0pt\a*-\a*\fi\ifdim\b*<0pt\b*-\b*\fi
\ifdim\a*>\b*\c*=.96\a*\advance\c*.4\b*
\else\c*=.96\b*\advance\c*.4\a*\fi
\k*\a*\multiply\k*\k*\l*\b*\multiply\l*\l*
\m*\k*\advance\m*\l*\n*\c*\r*\n*\multiply\n*\n*
\dn*\m*\advance\dn*-\n*\divide\dn*2\divide\dn*\r*
\advance\r*\dn*
\c*=\the\Nhalfperiods5sp\c*=#1\c*\ifdim\c*<0pt\c*-\c*\fi
\multiply\c*\r*\N*\c*\divide\N*10000}

\def\dashlin#1(#2,#3){\rlap{\calcnum*#1(#2,#3)\relax
\d**=#1\Lengthunit\ifdim\d**<0pt\d**-\d**\fi
\divide\N*2\multiply\N*2\advance\N*\*one
\divide\d**\N*\sm*\n*\*one\sh*(#2,#3){\sl*}\loop
\advance\n*\*one\sh*(#2,#3){\sm*}\advance\n*\*one
\sh*(#2,#3){\sl*}\ifnum\n*<\N*\repeat}}

\def\dashdotlin#1(#2,#3){\rlap{\calcnum*#1(#2,#3)\relax
\d**=#1\Lengthunit\ifdim\d**<0pt\d**-\d**\fi
\divide\N*2\multiply\N*2\advance\N*1\multiply\N*2\relax
\divide\d**\N*\sm*\n*\*two\sh*(#2,#3){\sl*}\loop
\advance\n*\*one\sh*(#2,#3){\kern-1.48pt\lower.5pt\hbox{\rm.}}\relax
\advance\n*\*one\sh*(#2,#3){\sm*}\advance\n*\*two
\sh*(#2,#3){\sl*}\ifnum\n*<\N*\repeat}}

\def\shl*(#1,#2)#3{\kern#1#3\lower#2#3\hbox{\unhcopy\spl*}}

\def\trianglin#1(#2,#3){\rlap{\toks0={#2}\toks1={#3}\calcnum*#1(#2,#3)\relax
\dd*=.57\Lengthunit\dd*=#1\dd*\divide\dd*\N*
\divide\dd*\*ths \multiply\dd*\magnitude
\d**=#1\Lengthunit\ifdim\d**<0pt\d**-\d**\fi
\multiply\N*2\divide\d**\N*\sm*\n*\*one\loop
\shl**{\dd*}\dd*-\dd*\advance\n*2\relax
\ifnum\n*<\N*\repeat\n*\N*\shl**{0pt}}}

\def\wavelin#1(#2,#3){\rlap{\toks0={#2}\toks1={#3}\calcnum*#1(#2,#3)\relax
\dd*=.23\Lengthunit\dd*=#1\dd*\divide\dd*\N*
\divide\dd*\*ths \multiply\dd*\magnitude
\d**=#1\Lengthunit\ifdim\d**<0pt\d**-\d**\fi
\multiply\N*4\divide\d**\N*\sm*\n*\*one\loop
\shl**{\dd*}\dt*=1.3\dd*\advance\n*\*one
\shl**{\dt*}\advance\n*\*one
\shl**{\dd*}\advance\n*\*two
\dd*-\dd*\ifnum\n*<\N*\repeat\n*\N*\shl**{0pt}}}

\def\w*lin(#1,#2){\rlap{\toks0={#1}\toks1={#2}\d**=\Lengthunit\dd*=-.12\d**
\divide\dd*\*ths \multiply\dd*\magnitude
\N*8\divide\d**\N*\sm*\n*\*one\loop
\shl**{\dd*}\dt*=1.3\dd*\advance\n*\*one
\shl**{\dt*}\advance\n*\*one
\shl**{\dd*}\advance\n*\*one
\shl**{0pt}\dd*-\dd*\advance\n*1\ifnum\n*<\N*\repeat}}

\def\l*arc(#1,#2)[#3][#4]{\rlap{\toks0={#1}\toks1={#2}\d**=\Lengthunit
\dd*=#3.037\d**\dd*=#4\dd*\dt*=#3.049\d**\dt*=#4\dt*\ifdim\d**>10mm\relax
\d**=.25\d**\n*\*one\shl**{-\dd*}\n*\*two\shl**{-\dt*}\n*3\relax
\shl**{-\dd*}\n*4\relax\shl**{0pt}\else
\ifdim\d**>5mm\d**=.5\d**\n*\*one\shl**{-\dt*}\n*\*two
\shl**{0pt}\else\n*\*one\shl**{0pt}\fi\fi}}

\def\d*arc(#1,#2)[#3][#4]{\rlap{\toks0={#1}\toks1={#2}\d**=\Lengthunit
\dd*=#3.037\d**\dd*=#4\dd*\d**=.25\d**\sm*\n*\*one\shl**{-\dd*}\relax
\n*3\relax\sh*(#1,#2){\xL*=\xscale\dd*\yL*=\yscale\dd*
\kern#2\xL*\lower#1\yL*\hbox{\sm*}}\n*4\relax\shl**{0pt}}}

\def\shl**#1{\c*=\the\n*\d**\d*=#1\relax
\a*=\the\toks0\c*\b*=\the\toks1\d*\advance\a*-\b*
\b*=\the\toks1\c*\d*=\the\toks0\d*\advance\b*\d*
\a*=\xscale\a*\b*=\yscale\b*
\rx* \the\cos*\a* \tmp* \the\sin*\b* \advance\rx*-\tmp*
\ry* \the\cos*\b* \tmp* \the\sin*\a* \advance\ry*\tmp*
\raise\ry*\rlap{\kern\rx*\unhcopy\spl*}}

\def\wlin*#1(#2,#3)[#4]{\rlap{\toks0={#2}\toks1={#3}\relax
\c*=#1\l*\c*\c*=.01\Lengthunit\m*\c*\divide\l*\m*
\c*=\the\Nhalfperiods5sp\multiply\c*\l*\N*\c*\divide\N*\*ths
\divide\N*2\multiply\N*2\advance\N*\*one
\dd*=.002\Lengthunit\dd*=#4\dd*\multiply\dd*\l*\divide\dd*\N*
\divide\dd*\*ths \multiply\dd*\magnitude
\d**=#1\multiply\N*4\divide\d**\N*\sm*\n*\*one\loop
\shl**{\dd*}\dt*=1.3\dd*\advance\n*\*one
\shl**{\dt*}\advance\n*\*one
\shl**{\dd*}\advance\n*\*two
\dd*-\dd*\ifnum\n*<\N*\repeat\n*\N*\shl**{0pt}}}

\def\wavebox#1{\setbox0\hbox{#1}\relax
\a*=\wd0\advance\a*14pt\b*=\ht0\advance\b*\dp0\advance\b*14pt\relax
\hbox{\kern9pt\relax
\rmov*(0pt,\ht0){\rmov*(-7pt,7pt){\wlin*\a*(1,0)[+]\wlin*\b*(0,-1)[-]}}\relax
\rmov*(\wd0,-\dp0){\rmov*(7pt,-7pt){\wlin*\a*(-1,0)[+]\wlin*\b*(0,1)[-]}}\relax
\box0\kern9pt}}

\def\rectangle#1(#2,#3){\relax
\lin#1(#2,0)\lin#1(0,#3)\mov#1(0,#3){\lin#1(#2,0)}\mov#1(#2,0){\lin#1(0,#3)}}

\def\dashrectangle#1(#2,#3){\dashlin#1(#2,0)\dashlin#1(0,#3)\relax
\mov#1(0,#3){\dashlin#1(#2,0)}\mov#1(#2,0){\dashlin#1(0,#3)}}

\def\waverectangle#1(#2,#3){\L*=#1\Lengthunit\a*=#2\L*\b*=#3\L*
\ifdim\a*<0pt\a*-\a*\def\x*{-1}\else\def\x*{1}\fi
\ifdim\b*<0pt\b*-\b*\def\y*{-1}\else\def\y*{1}\fi
\wlin*\a*(\x*,0)[-]\wlin*\b*(0,\y*)[+]\relax
\mov#1(0,#3){\wlin*\a*(\x*,0)[+]}\mov#1(#2,0){\wlin*\b*(0,\y*)[-]}}

\def\calcparab*{\ifnum\n*>\m*\k*\N*\advance\k*-\n*\else\k*\n*\fi
\a*=\the\k* sp\a*=10\a*\b*\dm*\advance\b*-\a*\k*\b*
\a*=\the\*ths\b*\divide\a*\l*\multiply\a*\k*
\divide\a*\l*\k*\*ths\r*\a*\advance\k*-\r*\dt*=\the\k*\L*}

\def\arcto#1(#2,#3)[#4]{\rlap{\toks0={#2}\toks1={#3}\calcnum*#1(#2,#3)\relax
\dm*=135sp\dm*=#1\dm*\d**=#1\Lengthunit\ifdim\dm*<0pt\dm*-\dm*\fi
\multiply\dm*\r*\a*=.3\dm*\a*=#4\a*\ifdim\a*<0pt\a*-\a*\fi
\advance\dm*\a*\N*\dm*\divide\N*10000\relax
\divide\N*2\multiply\N*2\advance\N*\*one
\L*=-.25\d**\L*=#4\L*\divide\d**\N*\divide\L*\*ths
\m*\N*\divide\m*2\dm*=\the\m*5sp\l*\dm*\sm*\n*\*one\loop
\calcparab*\shl**{-\dt*}\advance\n*1\ifnum\n*<\N*\repeat}}

\def\arrarcto#1(#2,#3)[#4]{\L*=#1\Lengthunit\L*=.54\L*
\arcto#1(#2,#3)[#4]\rmov*(#2\L*,#3\L*){\d*=.457\L*\d*=#4\d*\d**-\d*
\rmov*(#3\d**,#2\d*){\arrow.02(#2,#3)}}}

\def\dasharcto#1(#2,#3)[#4]{\rlap{\toks0={#2}\toks1={#3}\relax
\calcnum*#1(#2,#3)\dm*=\the\N*5sp\a*=.3\dm*\a*=#4\a*\ifdim\a*<0pt\a*-\a*\fi
\advance\dm*\a*\N*\dm*
\divide\N*20\multiply\N*2\advance\N*1\d**=#1\Lengthunit
\L*=-.25\d**\L*=#4\L*\divide\d**\N*\divide\L*\*ths
\m*\N*\divide\m*2\dm*=\the\m*5sp\l*\dm*
\sm*\n*\*one\loop\calcparab*
\shl**{-\dt*}\advance\n*1\ifnum\n*>\N*\else\calcparab*
\sh*(#2,#3){\xL*=#3\dt* \yL*=#2\dt*
\rx* \the\cos*\xL* \tmp* \the\sin*\yL* \advance\rx*\tmp*
\ry* \the\cos*\yL* \tmp* \the\sin*\xL* \advance\ry*-\tmp*
\kern\rx*\lower\ry*\hbox{\sm*}}\fi
\advance\n*1\ifnum\n*<\N*\repeat}}

\def\*shl*#1{\c*=\the\n*\d**\advance\c*#1\a**\d*\dt*\advance\d*#1\b**
\a*=\the\toks0\c*\b*=\the\toks1\d*\advance\a*-\b*
\b*=\the\toks1\c*\d*=\the\toks0\d*\advance\b*\d*
\rx* \the\cos*\a* \tmp* \the\sin*\b* \advance\rx*-\tmp*
\ry* \the\cos*\b* \tmp* \the\sin*\a* \advance\ry*\tmp*
\raise\ry*\rlap{\kern\rx*\unhcopy\spl*}}

\def\calcnormal*#1{\b**=10000sp\a**\b**\k*\n*\advance\k*-\m*
\multiply\a**\k*\divide\a**\m*\a**=#1\a**\ifdim\a**<0pt\a**-\a**\fi
\ifdim\a**>\b**\d*=.96\a**\advance\d*.4\b**
\else\d*=.96\b**\advance\d*.4\a**\fi
\d*=.01\d*\r*\d*\divide\a**\r*\divide\b**\r*
\ifnum\k*<0\a**-\a**\fi\d*=#1\d*\ifdim\d*<0pt\b**-\b**\fi
\k*\a**\a**=\the\k*\dd*\k*\b**\b**=\the\k*\dd*}

\def\wavearcto#1(#2,#3)[#4]{\rlap{\toks0={#2}\toks1={#3}\relax
\calcnum*#1(#2,#3)\c*=\the\N*5sp\a*=.4\c*\a*=#4\a*\ifdim\a*<0pt\a*-\a*\fi
\advance\c*\a*\N*\c*\divide\N*20\multiply\N*2\advance\N*-1\multiply\N*4\relax
\d**=#1\Lengthunit\dd*=.012\d**
\divide\dd*\*ths \multiply\dd*\magnitude
\ifdim\d**<0pt\d**-\d**\fi\L*=.25\d**
\divide\d**\N*\divide\dd*\N*\L*=#4\L*\divide\L*\*ths
\m*\N*\divide\m*2\dm*=\the\m*0sp\l*\dm*
\sm*\n*\*one\loop\calcnormal*{#4}\calcparab*
\*shl*{1}\advance\n*\*one\calcparab*
\*shl*{1.3}\advance\n*\*one\calcparab*
\*shl*{1}\advance\n*2\dd*-\dd*\ifnum\n*<\N*\repeat\n*\N*\shl**{0pt}}}

\def\triangarcto#1(#2,#3)[#4]{\rlap{\toks0={#2}\toks1={#3}\relax
\calcnum*#1(#2,#3)\c*=\the\N*5sp\a*=.4\c*\a*=#4\a*\ifdim\a*<0pt\a*-\a*\fi
\advance\c*\a*\N*\c*\divide\N*20\multiply\N*2\advance\N*-1\multiply\N*2\relax
\d**=#1\Lengthunit\dd*=.012\d**
\divide\dd*\*ths \multiply\dd*\magnitude
\ifdim\d**<0pt\d**-\d**\fi\L*=.25\d**
\divide\d**\N*\divide\dd*\N*\L*=#4\L*\divide\L*\*ths
\m*\N*\divide\m*2\dm*=\the\m*0sp\l*\dm*
\sm*\n*\*one\loop\calcnormal*{#4}\calcparab*
\*shl*{1}\advance\n*2\dd*-\dd*\ifnum\n*<\N*\repeat\n*\N*\shl**{0pt}}}

\def\hr*#1{\L*=\xscale\Lengthunit\ifnum
\angle**=0\clap{\vrule width#1\L* height.1pt}\else
\L*=#1\L*\L*=.5\L*\rmov*(-\L*,0pt){\sm*}\rmov*(\L*,0pt){\sl*}\fi}

\def\shade#1[#2]{\rlap{\Lengthunit=#1\Lengthunit
\special{em:linewidth .001pt}\relax
\mov(0,#2.05){\hr*{.994}}\mov(0,#2.1){\hr*{.980}}\relax
\mov(0,#2.15){\hr*{.953}}\mov(0,#2.2){\hr*{.916}}\relax
\mov(0,#2.25){\hr*{.867}}\mov(0,#2.3){\hr*{.798}}\relax
\mov(0,#2.35){\hr*{.715}}\mov(0,#2.4){\hr*{.603}}\relax
\mov(0,#2.45){\hr*{.435}}\special{em:linewidth \the\linwid*}}}

\def\dshade#1[#2]{\rlap{\special{em:linewidth .001pt}\relax
\Lengthunit=#1\Lengthunit\if#2-\def\t*{+}\else\def\t*{-}\fi
\mov(0,\t*.025){\relax
\mov(0,#2.05){\hr*{.995}}\mov(0,#2.1){\hr*{.988}}\relax
\mov(0,#2.15){\hr*{.969}}\mov(0,#2.2){\hr*{.937}}\relax
\mov(0,#2.25){\hr*{.893}}\mov(0,#2.3){\hr*{.836}}\relax
\mov(0,#2.35){\hr*{.760}}\mov(0,#2.4){\hr*{.662}}\relax
\mov(0,#2.45){\hr*{.531}}\mov(0,#2.5){\hr*{.320}}\relax
\special{em:linewidth \the\linwid*}}}}

\def\vdot{\rlap{\kern-1.9pt\lower1.8pt\hbox{$\scriptstyle\bullet$}}}
\def\vtimes{\rlap{\kern-3pt\lower1.8pt\hbox{$\scriptstyle\times$}}}
\def\vDot{\rlap{\kern-2.3pt\lower2.7pt\hbox{$\bullet$}}}
\def\vTimes{\rlap{\kern-3.6pt\lower2.4pt\hbox{$\times$}}}

\def\arc(#1)[#2,#3]{{\k*=#2\l*=#3\m*=\l*
\advance\m*-6\ifnum\k*>\l*\relax\else
{\rotate(#2)\mov(#1,0){\sm*}}\loop
\ifnum\k*<\m*\advance\k*5{\rotate(\k*)\mov(#1,0){\sl*}}\repeat
{\rotate(#3)\mov(#1,0){\sl*}}\fi}}

\def\dasharc(#1)[#2,#3]{{\k**=#2\n*=#3\advance\n*-1\advance\n*-\k**
\L*=1000sp\L*#1\L* \multiply\L*\n* \multiply\L*\Nhalfperiods
\divide\L*57\N*\L* \divide\N*2000\ifnum\N*=0\N*1\fi
\r*\n*	\divide\r*\N* \ifnum\r*<2\r*2\fi
\m**\r* \divide\m**2 \l**\r* \advance\l**-\m** \N*\n* \divide\N*\r*
\k**\r* \multiply\k**\N* \dn*\n* \advance\dn*-\k** \divide\dn*2\advance\dn*\*one
\r*\l** \divide\r*2\advance\dn*\r* \advance\N*-2\k**#2\relax
\ifnum\l**<6{\rotate(#2)\mov(#1,0){\sm*}}\advance\k**\dn*
{\rotate(\k**)\mov(#1,0){\sl*}}\advance\k**\m**
{\rotate(\k**)\mov(#1,0){\sm*}}\loop
\advance\k**\l**{\rotate(\k**)\mov(#1,0){\sl*}}\advance\k**\m**
{\rotate(\k**)\mov(#1,0){\sm*}}\advance\N*-1\ifnum\N*>0\repeat
{\rotate(#3)\mov(#1,0){\sl*}}\else\advance\k**\dn*
\arc(#1)[#2,\k**]\loop\advance\k**\m** \r*\k**
\advance\k**\l** {\arc(#1)[\r*,\k**]}\relax
\advance\N*-1\ifnum\N*>0\repeat
\advance\k**\m**\arc(#1)[\k**,#3]\fi}}

\def\triangarc#1(#2)[#3,#4]{{\k**=#3\n*=#4\advance\n*-\k**
\L*=1000sp\L*#2\L* \multiply\L*\n* \multiply\L*\Nhalfperiods
\divide\L*57\N*\L* \divide\N*1000\ifnum\N*=0\N*1\fi
\d**=#2\Lengthunit \d*\d** \divide\d*57\multiply\d*\n*
\r*\n*	\divide\r*\N* \ifnum\r*<2\r*2\fi
\m**\r* \divide\m**2 \l**\r* \advance\l**-\m** \N*\n* \divide\N*\r*
\dt*\d* \divide\dt*\N* \dt*.5\dt* \dt*#1\dt*
\divide\dt*1000\multiply\dt*\magnitude
\k**\r* \multiply\k**\N* \dn*\n* \advance\dn*-\k** \divide\dn*2\relax
\r*\l** \divide\r*2\advance\dn*\r* \advance\N*-1\k**#3\relax
{\rotate(#3)\mov(#2,0){\sm*}}\advance\k**\dn*
{\rotate(\k**)\mov(#2,0){\sl*}}\advance\k**-\m**\advance\l**\m**\loop\dt*-\dt*
\d*\d** \advance\d*\dt*
\advance\k**\l**{\rotate(\k**)\rmov*(\d*,0pt){\sl*}}%
\advance\N*-1\ifnum\N*>0\repeat\advance\k**\m**
{\rotate(\k**)\mov(#2,0){\sl*}}{\rotate(#4)\mov(#2,0){\sl*}}}}

\def\wavearc#1(#2)[#3,#4]{{\k**=#3\n*=#4\advance\n*-\k**
\L*=4000sp\L*#2\L* \multiply\L*\n* \multiply\L*\Nhalfperiods
\divide\L*57\N*\L* \divide\N*1000\ifnum\N*=0\N*1\fi
\d**=#2\Lengthunit \d*\d** \divide\d*57\multiply\d*\n*
\r*\n*	\divide\r*\N* \ifnum\r*=0\r*1\fi
\m**\r* \divide\m**2 \l**\r* \advance\l**-\m** \N*\n* \divide\N*\r*
\dt*\d* \divide\dt*\N* \dt*.7\dt* \dt*#1\dt*
\divide\dt*1000\multiply\dt*\magnitude
\k**\r* \multiply\k**\N* \dn*\n* \advance\dn*-\k** \divide\dn*2\relax
\divide\N*4\advance\N*-1\k**#3\relax
{\rotate(#3)\mov(#2,0){\sm*}}\advance\k**\dn*
{\rotate(\k**)\mov(#2,0){\sl*}}\advance\k**-\m**\advance\l**\m**\loop\dt*-\dt*
\d*\d** \advance\d*\dt* \dd*\d** \advance\dd*1.3\dt*
\advance\k**\r*{\rotate(\k**)\rmov*(\d*,0pt){\sl*}}\relax
\advance\k**\r*{\rotate(\k**)\rmov*(\dd*,0pt){\sl*}}\relax
\advance\k**\r*{\rotate(\k**)\rmov*(\d*,0pt){\sl*}}\relax
\advance\k**\r*
\advance\N*-1\ifnum\N*>0\repeat\advance\k**\m**
{\rotate(\k**)\mov(#2,0){\sl*}}{\rotate(#4)\mov(#2,0){\sl*}}}}

\def\gmov*#1(#2,#3)#4{\rlap{\L*=#1\Lengthunit
\xL*=#2\L* \yL*=#3\L*
\rx* \gcos*\xL* \tmp* \gsin*\yL* \advance\rx*-\tmp*
\ry* \gcos*\yL* \tmp* \gsin*\xL* \advance\ry*\tmp*
\rx*=\xscale\rx* \ry*=\yscale\ry*
\xL* \the\cos*\rx* \tmp* \the\sin*\ry* \advance\xL*-\tmp*
\yL* \the\cos*\ry* \tmp* \the\sin*\rx* \advance\yL*\tmp*
\kern\xL*\raise\yL*\hbox{#4}}}

\def\rgmov*(#1,#2)#3{\rlap{\xL*#1\yL*#2\relax
\rx* \gcos*\xL* \tmp* \gsin*\yL* \advance\rx*-\tmp*
\ry* \gcos*\yL* \tmp* \gsin*\xL* \advance\ry*\tmp*
\rx*=\xscale\rx* \ry*=\yscale\ry*
\xL* \the\cos*\rx* \tmp* \the\sin*\ry* \advance\xL*-\tmp*
\yL* \the\cos*\ry* \tmp* \the\sin*\rx* \advance\yL*\tmp*
\kern\xL*\raise\yL*\hbox{#3}}}

\def\Earc(#1)[#2,#3][#4,#5]{{\k*=#2\l*=#3\m*=\l*
\advance\m*-6\ifnum\k*>\l*\relax\else\def\xscale{#4}\def\yscale{#5}\relax
{\angle**0\rotate(#2)}\gmov*(#1,0){\sm*}\loop
\ifnum\k*<\m*\advance\k*5\relax
{\angle**0\rotate(\k*)}\gmov*(#1,0){\sl*}\repeat
{\angle**0\rotate(#3)}\gmov*(#1,0){\sl*}\relax
\def\xscale{1}\def\yscale{1}\fi}}

\def\dashEarc(#1)[#2,#3][#4,#5]{{\k**=#2\n*=#3\advance\n*-1\advance\n*-\k**
\L*=1000sp\L*#1\L* \multiply\L*\n* \multiply\L*\Nhalfperiods
\divide\L*57\N*\L* \divide\N*2000\ifnum\N*=0\N*1\fi
\r*\n*	\divide\r*\N* \ifnum\r*<2\r*2\fi
\m**\r* \divide\m**2 \l**\r* \advance\l**-\m** \N*\n* \divide\N*\r*
\k**\r*\multiply\k**\N* \dn*\n* \advance\dn*-\k** \divide\dn*2\advance\dn*\*one
\r*\l** \divide\r*2\advance\dn*\r* \advance\N*-2\k**#2\relax
\ifnum\l**<6\def\xscale{#4}\def\yscale{#5}\relax
{\angle**0\rotate(#2)}\gmov*(#1,0){\sm*}\advance\k**\dn*
{\angle**0\rotate(\k**)}\gmov*(#1,0){\sl*}\advance\k**\m**
{\angle**0\rotate(\k**)}\gmov*(#1,0){\sm*}\loop
\advance\k**\l**{\angle**0\rotate(\k**)}\gmov*(#1,0){\sl*}\advance\k**\m**
{\angle**0\rotate(\k**)}\gmov*(#1,0){\sm*}\advance\N*-1\ifnum\N*>0\repeat
{\angle**0\rotate(#3)}\gmov*(#1,0){\sl*}\def\xscale{1}\def\yscale{1}\else
\advance\k**\dn* \Earc(#1)[#2,\k**][#4,#5]\loop\advance\k**\m** \r*\k**
\advance\k**\l** {\Earc(#1)[\r*,\k**][#4,#5]}\relax
\advance\N*-1\ifnum\N*>0\repeat
\advance\k**\m**\Earc(#1)[\k**,#3][#4,#5]\fi}}

\def\triangEarc#1(#2)[#3,#4][#5,#6]{{\k**=#3\n*=#4\advance\n*-\k**
\L*=1000sp\L*#2\L* \multiply\L*\n* \multiply\L*\Nhalfperiods
\divide\L*57\N*\L* \divide\N*1000\ifnum\N*=0\N*1\fi
\d**=#2\Lengthunit \d*\d** \divide\d*57\multiply\d*\n*
\r*\n*	\divide\r*\N* \ifnum\r*<2\r*2\fi
\m**\r* \divide\m**2 \l**\r* \advance\l**-\m** \N*\n* \divide\N*\r*
\dt*\d* \divide\dt*\N* \dt*.5\dt* \dt*#1\dt*
\divide\dt*1000\multiply\dt*\magnitude
\k**\r* \multiply\k**\N* \dn*\n* \advance\dn*-\k** \divide\dn*2\relax
\r*\l** \divide\r*2\advance\dn*\r* \advance\N*-1\k**#3\relax
\def\xscale{#5}\def\yscale{#6}\relax
{\angle**0\rotate(#3)}\gmov*(#2,0){\sm*}\advance\k**\dn*
{\angle**0\rotate(\k**)}\gmov*(#2,0){\sl*}\advance\k**-\m**
\advance\l**\m**\loop\dt*-\dt* \d*\d** \advance\d*\dt*
\advance\k**\l**{\angle**0\rotate(\k**)}\rgmov*(\d*,0pt){\sl*}\relax
\advance\N*-1\ifnum\N*>0\repeat\advance\k**\m**
{\angle**0\rotate(\k**)}\gmov*(#2,0){\sl*}\relax
{\angle**0\rotate(#4)}\gmov*(#2,0){\sl*}\def\xscale{1}\def\yscale{1}}}

\def\waveEarc#1(#2)[#3,#4][#5,#6]{{\k**=#3\n*=#4\advance\n*-\k**
\L*=4000sp\L*#2\L* \multiply\L*\n* \multiply\L*\Nhalfperiods
\divide\L*57\N*\L* \divide\N*1000\ifnum\N*=0\N*1\fi
\d**=#2\Lengthunit \d*\d** \divide\d*57\multiply\d*\n*
\r*\n*	\divide\r*\N* \ifnum\r*=0\r*1\fi
\m**\r* \divide\m**2 \l**\r* \advance\l**-\m** \N*\n* \divide\N*\r*
\dt*\d* \divide\dt*\N* \dt*.7\dt* \dt*#1\dt*
\divide\dt*1000\multiply\dt*\magnitude
\k**\r* \multiply\k**\N* \dn*\n* \advance\dn*-\k** \divide\dn*2\relax
\divide\N*4\advance\N*-1\k**#3\def\xscale{#5}\def\yscale{#6}\relax
{\angle**0\rotate(#3)}\gmov*(#2,0){\sm*}\advance\k**\dn*
{\angle**0\rotate(\k**)}\gmov*(#2,0){\sl*}\advance\k**-\m**
\advance\l**\m**\loop\dt*-\dt*
\d*\d** \advance\d*\dt* \dd*\d** \advance\dd*1.3\dt*
\advance\k**\r*{\angle**0\rotate(\k**)}\rgmov*(\d*,0pt){\sl*}\relax
\advance\k**\r*{\angle**0\rotate(\k**)}\rgmov*(\dd*,0pt){\sl*}\relax
\advance\k**\r*{\angle**0\rotate(\k**)}\rgmov*(\d*,0pt){\sl*}\relax
\advance\k**\r*
\advance\N*-1\ifnum\N*>0\repeat\advance\k**\m**
{\angle**0\rotate(\k**)}\gmov*(#2,0){\sl*}\relax
{\angle**0\rotate(#4)}\gmov*(#2,0){\sl*}\def\xscale{1}\def\yscale{1}}}

\newcount\CatcodeOfAtSign
\CatcodeOfAtSign=\the\catcode`\@
\catcode`\@=11
\def\@arc#1[#2][#3]{\rlap{\Lengthunit=#1\Lengthunit
\sm*\l*arc(#2.1914,#3.0381)[#2][#3]\relax
\mov(#2.1914,#3.0381){\l*arc(#2.1622,#3.1084)[#2][#3]}\relax
\mov(#2.3536,#3.1465){\l*arc(#2.1084,#3.1622)[#2][#3]}\relax
\mov(#2.4619,#3.3086){\l*arc(#2.0381,#3.1914)[#2][#3]}}}

\def\dash@arc#1[#2][#3]{\rlap{\Lengthunit=#1\Lengthunit
\d*arc(#2.1914,#3.0381)[#2][#3]\relax
\mov(#2.1914,#3.0381){\d*arc(#2.1622,#3.1084)[#2][#3]}\relax
\mov(#2.3536,#3.1465){\d*arc(#2.1084,#3.1622)[#2][#3]}\relax
\mov(#2.4619,#3.3086){\d*arc(#2.0381,#3.1914)[#2][#3]}}}

\def\wave@arc#1[#2][#3]{\rlap{\Lengthunit=#1\Lengthunit
\w*lin(#2.1914,#3.0381)\relax
\mov(#2.1914,#3.0381){\w*lin(#2.1622,#3.1084)}\relax
\mov(#2.3536,#3.1465){\w*lin(#2.1084,#3.1622)}\relax
\mov(#2.4619,#3.3086){\w*lin(#2.0381,#3.1914)}}}

\def\bezier#1(#2,#3)(#4,#5)(#6,#7){\N*#1\l*\N* \advance\l*\*one
\d* #4\Lengthunit \advance\d* -#2\Lengthunit \multiply\d* \*two
\b* #6\Lengthunit \advance\b* -#2\Lengthunit
\advance\b*-\d* \divide\b*\N*
\d** #5\Lengthunit \advance\d** -#3\Lengthunit \multiply\d** \*two
\b** #7\Lengthunit \advance\b** -#3\Lengthunit
\advance\b** -\d** \divide\b**\N*
\mov(#2,#3){\sm*{\loop\ifnum\m*<\l*
\a*\m*\b* \advance\a*\d* \divide\a*\N* \multiply\a*\m*
\a**\m*\b** \advance\a**\d** \divide\a**\N* \multiply\a**\m*
\rmov*(\a*,\a**){\unhcopy\spl*}\advance\m*\*one\repeat}}}

\catcode`\*=12

\newcount\n@ast
\def\n@ast@#1{\n@ast0\relax\get@ast@#1\end}
\def\get@ast@#1{\ifx#1\end\let\next\relax\else
\ifx#1*\advance\n@ast1\fi\let\next\get@ast@\fi\next}

\newif\if@up \newif\if@dwn
\def\up@down@#1{\@upfalse\@dwnfalse
\if#1u\@uptrue\fi\if#1U\@uptrue\fi\if#1+\@uptrue\fi
\if#1d\@dwntrue\fi\if#1D\@dwntrue\fi\if#1-\@dwntrue\fi}

\def\halfcirc#1(#2)[#3]{{\Lengthunit=#2\Lengthunit\up@down@{#3}\relax
\if@up\mov(0,.5){\@arc[-][-]\@arc[+][-]}\fi
\if@dwn\mov(0,-.5){\@arc[-][+]\@arc[+][+]}\fi
\def\lft{\mov(0,.5){\@arc[-][-]}\mov(0,-.5){\@arc[-][+]}}\relax
\def\rght{\mov(0,.5){\@arc[+][-]}\mov(0,-.5){\@arc[+][+]}}\relax
\if#3l\lft\fi\if#3L\lft\fi\if#3r\rght\fi\if#3R\rght\fi
\n@ast@{#1}\relax
\ifnum\n@ast>0\if@up\shade[+]\fi\if@dwn\shade[-]\fi\fi
\ifnum\n@ast>1\if@up\dshade[+]\fi\if@dwn\dshade[-]\fi\fi}}

\def\halfdashcirc(#1)[#2]{{\Lengthunit=#1\Lengthunit\up@down@{#2}\relax
\if@up\mov(0,.5){\dash@arc[-][-]\dash@arc[+][-]}\fi
\if@dwn\mov(0,-.5){\dash@arc[-][+]\dash@arc[+][+]}\fi
\def\lft{\mov(0,.5){\dash@arc[-][-]}\mov(0,-.5){\dash@arc[-][+]}}\relax
\def\rght{\mov(0,.5){\dash@arc[+][-]}\mov(0,-.5){\dash@arc[+][+]}}\relax
\if#2l\lft\fi\if#2L\lft\fi\if#2r\rght\fi\if#2R\rght\fi}}

\def\halfwavecirc(#1)[#2]{{\Lengthunit=#1\Lengthunit\up@down@{#2}\relax
\if@up\mov(0,.5){\wave@arc[-][-]\wave@arc[+][-]}\fi
\if@dwn\mov(0,-.5){\wave@arc[-][+]\wave@arc[+][+]}\fi
\def\lft{\mov(0,.5){\wave@arc[-][-]}\mov(0,-.5){\wave@arc[-][+]}}\relax
\def\rght{\mov(0,.5){\wave@arc[+][-]}\mov(0,-.5){\wave@arc[+][+]}}\relax
\if#2l\lft\fi\if#2L\lft\fi\if#2r\rght\fi\if#2R\rght\fi}}

\catcode`\*=11

\def\Circle#1(#2){\halfcirc#1(#2)[u]\halfcirc#1(#2)[d]\n@ast@{#1}\relax
\ifnum\n@ast>0\L*=\xscale\Lengthunit
\ifnum\angle**=0\clap{\vrule width#2\L* height.1pt}\else
\L*=#2\L*\L*=.5\L*\special{em:linewidth .001pt}\relax
\rmov*(-\L*,0pt){\sm*}\rmov*(\L*,0pt){\sl*}\relax
\special{em:linewidth \the\linwid*}\fi\fi}

\catcode`\*=12

\def\wavecirc(#1){\halfwavecirc(#1)[u]\halfwavecirc(#1)[d]}

\def\dashcirc(#1){\halfdashcirc(#1)[u]\halfdashcirc(#1)[d]}

\def\xscale{1}
\def\yscale{1}

\def\Ellipse#1(#2)[#3,#4]{\def\xscale{#3}\def\yscale{#4}\relax
\Circle#1(#2)\def\xscale{1}\def\yscale{1}}

\def\dashEllipse(#1)[#2,#3]{\def\xscale{#2}\def\yscale{#3}\relax
\dashcirc(#1)\def\xscale{1}\def\yscale{1}}

\def\waveEllipse(#1)[#2,#3]{\def\xscale{#2}\def\yscale{#3}\relax
\wavecirc(#1)\def\xscale{1}\def\yscale{1}}

\def\halfEllipse#1(#2)[#3][#4,#5]{\def\xscale{#4}\def\yscale{#5}\relax
\halfcirc#1(#2)[#3]\def\xscale{1}\def\yscale{1}}

\def\halfdashEllipse(#1)[#2][#3,#4]{\def\xscale{#3}\def\yscale{#4}\relax
\halfdashcirc(#1)[#2]\def\xscale{1}\def\yscale{1}}

\def\halfwaveEllipse(#1)[#2][#3,#4]{\def\xscale{#3}\def\yscale{#4}\relax
\halfwavecirc(#1)[#2]\def\xscale{1}\def\yscale{1}}

\catcode`\@=\the\CatcodeOfAtSign  

\begin{abstract}
The proton structure and proton polarizability corrections to the Lamb
shift of electronic hydrogen and muonic hydrogen were evaluated on the
basis of modern experimental data on deep inelastic structure functions.
Numerical value of proton polarizability contribution to (2P-2S) Lamb shift
is equal to 4.4 GHz.
\end{abstract}

\newpage

The muonic hydrogen $(\mu^- p^+)$ is a two-particle bound state of a muon
and a proton. The energy levels of this system are defined as for electronic
hydrogen \cite{SY}. By virtue of the fact that the mass ratio of electron and
muon $m_e/m_\mu = 4.836332\cdot 10^{-3}$, some quantum electrodynamical
contributions, proton structure corrections and proton polarizability effects
increase in importance in the case of muonic hydrogen \cite{Gi,BR}. For example,
the electron vacuum polarization gives the main contribution to the Lamb shift
of muonic hydrogen (contrary to ordinary hydrogen), because the Compton
wavelength of the electron and the Bohr radius of muonic hydrogen have the
same order: $\hbar/m_ec : \hbar^2/\mu e^2 = 0.737386 $.
The experimental measurement of (2P-2S) Lamb shift in muonic hydrogen
makes it possible to determine with higher accuracy the proton charge
radius $R_p = \sqrt{<r^2>}$ \cite{K}, which is one of the universal fundamental physical
constants. Theoretical investigations of the main contributions to the Lamb
shift of muonic hydrogen $(\mu p)$ were done in \cite{Gi,BR}. The results of
calculation of different order corrections, known at present time may be found
in \cite{P}. New six-order vacuum polarization contribution to the Lamb shift
of the muonic hydrogen was obtained recently in \cite{KN}. In this work we
consider the Lamb shift contribution in $(\mu^- p^+)$, connected with
the proton polarizability effects. As in the case of muonic hydrogen
hyperfine splitting \cite{MF1,MSF}, we have used the local quasipotential
equation for description of two-particle bound state $(\mu^- p^+)$ \cite{MF2}.

The proton polarizability contribution is determined by the amplitude of two -
photon muon - proton interaction, shown on Fig.1. Compton forward scattering
amplitude of the electron, appearing in the two - photon interaction, may be
expressed as a sum of direct and crossed two - photon diagrams:
\begin{equation}
M_{\mu\nu}^{(e)}=\bar u(q_1)\left[\gamma_\mu\frac{\hat{p_1}+\hat k+m_1}
{(p_1+k)^2-m_1^2+i\epsilon}\gamma_\nu+\gamma_\nu\frac{\hat{p_1}-\hat k+m_1}
{(p_1-k)^2-m_1^2+i\epsilon}\gamma_\mu\right]u(p_1).
\end{equation}

\begin{figure}
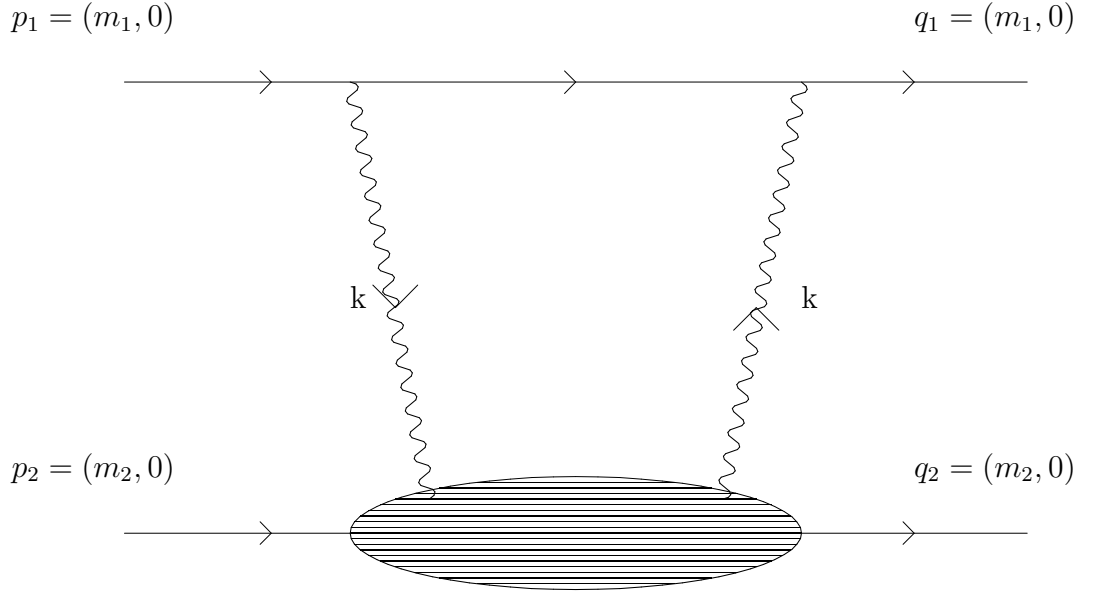

\magnitude=2000
\GRAPH(hsize=15){
\mov(2,0){\lin(2,0)}%
\mov(8,0){\lin(2,0)}%
\mov(2,4){\lin(8,0)}%
\mov(9,0){\lin(-0.1,-0.1)}%
\mov(9,0){\lin(-0.1,0.1)}%
\mov(9,4){\lin(-0.1,0.1)}%
\mov(9,4){\lin(-0.1,-0.1)}%
\mov(6,4){\lin(-0.1,-0.1)}%
\mov(6,4){\lin(-0.1,0.1)}%
\mov(4.4,2){\lin(-0.2,0.2)}%
\mov(4.4,2){\lin(0.2,0.2)}%
\mov(7.6,2){\lin(-0.2,-0.2)}%
\mov(7.6,2){\lin(0.2,-0.2)}%
\mov(4,2){k}%
\mov(8,2){k}%
\mov(3.3,0){\lin(-0.1,-0.1)}%
\mov(3.3,0){\lin(-0.1,0.1)}%
\mov(3.3,4){\lin(-0.1,0.1)}%
\mov(3.3,4){\lin(-0.1,-0.1)}%
\mov(6,0){\Ellipse*(1)[4,1]}%
\mov(4,4){\wavelin(0.7,-3.7)}%
\mov(8,4){\wavelin(-0.7,-3.7)}%
\mov(1,0.5){$p_2=(m_2,0)$}%
\mov(1,4.5){$p_1=(m_1,0)$}%
\mov(9,0.5){$q_2=(m_2,0)$}%
\mov(9,4.5){$q_1=(m_1,0)$}%
}
\vspace{5mm}
\caption{Proton polarizability correction}
\end{figure}

Neglecting by the relative motion momenta of the particles in initial and
final states ($\vec p=\vec q=0$), we can also parameterize the amplitude of
the proton virtual Compton scattering in the following manner \cite{B,Z}:
\begin{equation}
M_{\mu\nu}^{(p)}=\bar v(p_2)\Biggl\{\frac{1}{2}C_1\left(-g_{\mu\nu}+\frac{k_\mu k_\nu}
{k^2}\right)+\frac{1}{2m^2_2}C_2\left(p_{2 \mu}-\frac{m_2\nu}{k^2}k_\mu\right)
\left(p_{2 \nu}-\frac{m_2\nu}{k^2}k_\nu\right)+
\end{equation}
\begin{displaymath}
+\frac{1}{2m_2}H_1\left([\gamma_\nu,\hat k]p_{2 \mu}-[\gamma_\mu,\hat k]p_{2 nu}+[\gamma_\mu,
\gamma_\nu]\right)+
\end{displaymath}
\begin{displaymath}
+\frac{1}{2}H_2\left([\gamma_\nu,\hat k]k_\mu-[\gamma_\mu,\hat k]k_\nu+
[\gamma_\mu,\gamma_\nu]k^2\right)\Biggr\}v(q_2),
\end{displaymath}
where $\nu=k_0$ is the energy of the virtual photon. Tensor (2) contains
symmetric and antisymmetric parts. Symmetric part, which is determined by the
structure functions $C_{1,2}(\nu, k^2)$ fixes the contribution
to the Lamb shift. Antisymmetric part of expression (2), defined by the
functions $H_{1,2}(\nu, k^2)$, gives the contribution to the hyperfine
splitting of S - energy levels. To extract the hyperfine splitting part of
the quasipotential $V^{\rm hfs}$, we can use the projection operators of muon -
proton two - particle system on the states with total spin S=0 ($V(^1S_0)$)
and S=1 ($V(^3S_1)$):
\begin{equation}
\hat\Pi=u(p_1)\bar v(p_2)=\frac{1}{2\sqrt{2}}(1+\gamma_0)\hat\epsilon (\gamma_5),
\end{equation}
To construct the S -wave Lamb shift part of the quasipotential it is common practice
to use two approaches:

1. The averaging of the amplitudes (1) and (2) over muon and proton spins
\cite{BG}.

2. The use of projection operators (3) and the construction of necessary
interaction in the form:
\begin{equation}
V^{\rm Ls}=V(^1S_0)+\frac{3}{4}V^{\rm hfs}.
\end{equation}  

Multiplying the amplitudes (1) and (2) and extracting the corresponding
parts of the quasipotential (hyperfine splitting part and Lamb shift part),
we obtain:
\begin{equation}
\left(M_{\mu\nu}^{(\mu)}M_{\mu\nu}^{(p)}\right)^{\rm hfs}=-\frac{16}{3}\frac{k^2}
{k^4-4k_0^2m_1^2}\left[m_2(k_0^2+2k^2)H_1+3k_0k^2H_2\right],
\end{equation}
\begin{equation}
\left(M_{\mu\nu}^{(\mu)}M_{\mu\nu}^{(p)}\right)^{\rm Ls}=\frac{2m_1}
{k^4-4k_0^2m_1^2}\left[-(2k_0^2+k^2)C_1+(k^2-k_0^2)C_2\right],
\end{equation}

Expression (5) was used earlier in the hyperfine splitting calculation
\cite{DS,G,F}. Let us consider Lamb shift contribution on the basis of (6).
Using dispersion relation for structure functions $C_i(k_0, k^2)$ and
explicitly displaying the contribution of the proton intermediate state,
shown in Fig.2, we have:
\begin{equation}
C_i(k_0,k^2)=C_i^p(k_0,k^2)+\frac{1}{\pi}\int_{\nu_0}^\infty\frac{d\nu^2}
{(\nu^2-k_0^2)}Im C_i(\nu, k^2),
\end{equation}
\begin{displaymath}
\nu_0=m_\pi+\frac{1}{2m_2}(Q^2+m_\pi^2),~~Q^2=-k^2,
\end{displaymath}
and moreover the function $C_1(k_0,k^2)$ satisfies to dispersion relation
with one subtraction to provide the correct asymptotic behavior in $\nu$ of the intgral
expression in (7).
\begin{figure}
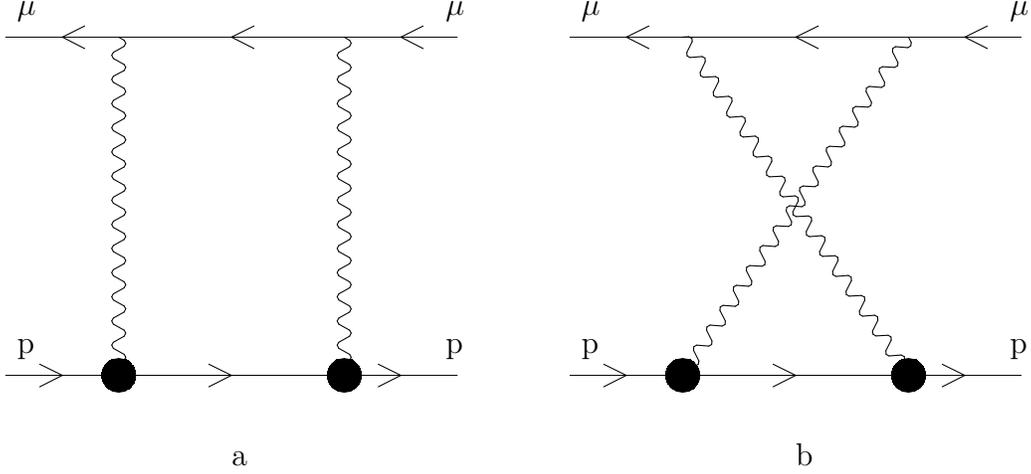

\magnitude=2000
\GRAPH(hsize=15){
\mov(0,0){\lin(1,0)}%
\mov(3,0){\lin(1,0)}%
\mov(0,3){\lin(4,0)}%
\mov(5,0){\lin(1,0)}%
\mov(8,0){\lin(1,0)}%
\mov(5,3){\lin(4,0)}%
\mov(3.,0.){\Circle**(0.3)}%
\mov(6.,0){\Circle**(0.3)}%
\mov(1.,0.){\Circle**(0.3)}%
\mov(8.,0.){\Circle**(0.3)}%
\mov(2.,-0.8){a}%
\mov(7.,-0.8){b}%  
\mov(0.5,3.){\lin(0.2,0.1)}%
\mov(0.5,3.){\lin(0.2,-0.1)}%
\mov(2.,3.){\lin(0.2,0.1)}%
\mov(2.,3.){\lin(0.2,-0.1)}%
\mov(3.5,3.){\lin(0.2,0.1)}%
\mov(3.5,3.){\lin(0.2,-0.1)}%  
\mov(5.5,3.){\lin(0.2,0.1)}%
\mov(5.5,3.){\lin(0.2,-0.1)}%
\mov(7.,3.){\lin(0.2,0.1)}%
\mov(7.,3.){\lin(0.2,-0.1)}%
\mov(8.5,3.){\lin(0.2,0.1)}%
\mov(8.5,3.){\lin(0.2,-0.1)}%  
\mov(0.5,0.){\lin(-0.2,0.1)}%
\mov(0.5,0.){\lin(-0.2,-0.1)}%
\mov(2.,0.){\lin(-0.2,0.1)}%
\mov(2.,0.){\lin(-0.2,-0.1)}%
\mov(3.5,0.){\lin(-0.2,0.1)}%
\mov(3.5,0.){\lin(-0.2,-0.1)}%  
\mov(5.5,0.){\lin(-0.2,0.1)}%
\mov(5.5,0.){\lin(-0.2,-0.1)}%
\mov(7.,0.){\lin(-0.2,0.1)}%
\mov(7.,0.){\lin(-0.2,-0.1)}%
\mov(8.5,0.){\lin(-0.2,0.1)}%
\mov(8.5,0.){\lin(-0.2,-0.1)}%  
\mov(1.,0){\lin(2.,0)}%
\mov(6.,0){\lin(2.,0)}%
%\mov(1.,-0.05){\rectangle(2,0.1)}%
%\mov(6.,-0.05){\rectangle(2,0.1)}%
\mov(1,0){\wavelin(0,3)}%
\mov(3,0){\wavelin(0,3)}%  
\mov(6,0){\wavelin(2,3)}%  
\mov(8,0){\wavelin(-2,3)}%
\mov(0.1,0.2){p}%
\mov(0.1,3.2){$\mu$}%
\mov(3.9,0.2){p}%
\mov(3.9,3.2){$\mu$}%
\mov(5.1,0.2){p}%
\mov(5.1,3.2){$\mu$}%
\mov(8.9,0.2){p}%
\mov(8.9,3.2){$\mu$}%
}
\vspace{5mm}

\caption{Proton finite - size correction of order $(Z\alpha)^5$.}
\end{figure}

The proton vertices of two - photon amplitudes are defined by two form
factors $F_1$ and $F_2$ \cite{BY}:
\begin{equation}
\Gamma^\mu=\gamma^\mu F_1+\frac{i}{2m_2}\sigma^{\mu\nu}k_\nu F_2,
\end{equation}
The contribution of diagrams (a) and (b) of Fig.2 to the Lamb shift may be
expressed as follows:
\begin{equation}
\Delta E^{\rm Ls}_p=-|\psi(0)|^2\frac{(Z\alpha)^2}{\pi^2}\int d^4k\frac{1}{(k^2)^2}
\frac{1}{2m_1}{m_2(k^4+4m_1^2k_0^2)}\frac{1}{(k^4+4m_2^2k_0^2)}
\end{equation}
\begin{displaymath}
\left[4m_2^2(k^4+2k_0^4)F_1^2-2k^4(k^2+2k_0^2)F_1F_2-3k_0^2k^4F_2^2\right],
\end{displaymath}
where we have transformed integration over four - dimensional Euclidian space,
changing $k_0\rightarrow ik_0$. It is necessary to say that this integration
contains infrared divergence. To remove it we subtract from (8) the correction
related with iteration term of the quasipotential and the contribution
of two - photon amplitudes in the case of point - like proton. The contribution
of iteration part of the quasipotential takes the form:
\begin{equation}
\Delta E^{\rm Ls}_{\rm iter}=V_{1\gamma}\times G^f\times V_{1\gamma}\approx
\frac{\mu^4(Z\alpha)^5}{m_2^2\pi n^3}\int_0^\infty\frac{dk}{k^2}\left[F_1'(0)
4m_2^2+F_2(0)\right].
\end{equation}
The iteration terms of quasipotential fully eliminate infrared divergences
in the hyperfine structure of hydrogen - like atoms \cite{Z}. Note that
the correction of order $(Z\alpha)^4$ in the Lamb shift of hydrogen atom,
depending on proton structure is defined by one - photon term of the
quasipotential $V_{1\gamma}\sim G_E(k^2)/\vec k^2$
\cite{RNF} ($G_E$ is the Sachs form factor) in the following form:
\begin{equation}
E^{\rm Ls}=\frac{2\mu^3}{3n^3}(Z\alpha)^4R^2_p\delta_{l0},~~R^2_p=\frac{1}{6G_E(0)}
\frac{\partial G_E(k^2)}{\partial k^2}|_{k^2=0},
\end{equation}
where $R_p$ is the proton charge radius \cite{K}. For point - like proton
the recoil correction to energy levels of order $(Z\alpha)^5$ is
determined by both non - relativistic and relativistic electron momenta
in the two - photon interaction operator $V_{2\gamma}$. The corresponding
expression takes the form \cite{SY}:
\begin{equation}
\Delta E_{\rm nl}=\frac{\mu^3}{m_1m_2}\frac{(Z\alpha)^5}{\pi n^3}\Biggl[\frac{2}{3}
\delta_{l0}\ln\left(\frac{1}{Z\alpha}\right)-\frac{8}{3}\ln k_0(n,l)-
\end{equation}
\begin{displaymath}
-\frac{1}{9}\delta_{l0}-\frac{7}{3}a_n-\frac{2}{m_2^2-m_1^2}\delta_{l0}
\left(m_2^2\ln\frac{m_2}{\mu}-m_1^2\ln\frac{m_2}{\mu}\right)\Biggr],
\end{displaymath}
\begin{equation}
a_n=-2\left[\ln\frac{2}{n}+\left(1+\frac{1}{2}+...+\frac{1}{n}\right)+1-\frac{1}{2n}
\right]\delta_{l0}+\frac{1-\delta_{l0}}{l(l+1)(2l+1)}.
\end{equation}

Taking into account the results (10), (12), we can express infrared finite contribution
of two - photon diagrams in the Lamb shift of hydrogen - like atom as follows:
\begin{equation}
\Delta E^{\rm Ls}_{2\gamma}=-\frac{\mu^3}{\pi n^3}\delta_{l0}(Z\alpha)^5\int_0^\infty
\frac{dk}{k}V(k),
\end{equation}
\begin{equation}
V(k)=\frac{2(F_1^2-1)}{m_1m_2}+\frac{8m_1[F_2(0)+4m_2^2F_1'(0)]}{m_2(m_1+m_2)k}+
\end{equation}
\begin{displaymath}
+\frac{k^2}{2m_1^3m_2^3}\left[2(F_1^2-1)(m_1^2+m_2^2)+4F_1F_2m_1^2+3F_2^2m_1^2\right]+
\end{displaymath}
\begin{displaymath}
+\frac{\sqrt{k^2+4m_1^2}}{2m_1^2m_2(m_1^2-m_2^2)k}\Biggl\{k^2\left[2(F_1^2-1)m_2^2+
4F_1F_2m_1^2+3F_2^2m_1^2\right]+
\end{displaymath}
\begin{displaymath}  
+8m_1^4F_1F_2+\frac{16m_1^4m_2^2(F_1^2-1)}{k^2}\Biggr\}-
\end{displaymath}
\begin{displaymath}
-\frac{\sqrt{k^2+4m_1^2}m_1}{2m_2^3(m_1^2-m_2^2)k}\Biggl\{k^2\left[2(F_1^2-1)m_2^2+
4F_1F_2+3F_2^2\right]-
\end{displaymath}
\begin{displaymath}
-8m_2^2F_1F_2+\frac{16m_2^4(F_1^2-1)}{k^2}\Biggr\}.
\end{displaymath}
To calculate the correction (14), we have used the dipole parameterization
of proton form factors:
\begin{equation}
F_1(k^2)=\frac{G_E-\frac{k^2}{4m_2^2}G_M}{1-\frac{k^2}{4m_2^2}},~~
F_2(k^2)=\frac{G_M-G_E}{1-\frac{k^2}{4m_2^2}},
\end{equation}
\begin{displaymath}
G_M=\frac{1+\kappa}{\left(1-\frac{k^2}{\Lambda^2}\right)^2},~~
G_E=\frac{1}{\left(1-\frac{k^2}{\Lambda^2}\right)^2},
\end{displaymath}
where proton structure parameter was taken as for calculation of hyperfine
splitting in hydrogen atom $\Lambda=0.898 m_2$ \cite{BY},
$\kappa$ =1.792847 is the proton anomalous magnetic moment.
Numerical value of correction (14) in the Lamb shift of electronic hydrogen
$\Delta E_H^{\rm Ls}$ (2P-2S) = 4.25 Hz and of muonic hydrogen
$\Delta E^{\rm Ls}_{\mu p} = 4.35 $ GHz, which coincides with the result of
\cite{P}. Let us consider now the effects of proton polarizability, which are
described by the dispersion integral (7). The imaginary parts of amplitudes
$C_i(k_0,k^2)$ are related to deep inelastic structure functions
$F_i(x,Q^2)$:
\begin{equation}
\frac{1}{\pi}Im C_1(x,Q^2)=\frac{F_1(x,Q^2)}{m_2},~~~\frac{1}{\pi}Im C_2(x,Q^2)=\frac{F_2(x,Q^2)}{\nu},~~x=\frac{Q^2}{2m_2\nu}.
\end{equation}
In the limit $Q^2\rightarrow 0$ the following conditions should be fulfilled:
\begin{equation}
F_2=O(Q^2),~~~\frac{F_1}{m_2}-\frac{F_2\nu}{Q^2}=O(Q^2).
\end{equation}
They play important role for the parameterization of structure functions
$F_i(x,Q^2)$ at low $Q^2$ \cite{BK}. Using relations (6) - (7) and
transforming the integration over four - dimensional Euclidian space in
the loop amplitudes
\begin{equation}
\int d^4k=4\pi\int_0^\infty k^3 dk\int_0^\pi Sin^2\phi d\phi,~~~k^0=k Cos\phi,
\end{equation}
we can represent the proton polarizability contribution to the Lamb shift
of muonic hydrogen in the following way:
\begin{equation}
\Delta E^{\rm Ls}_{\rm pol}=-\frac{16\mu^3(Z\alpha)^5m_1}{\pi^2 n^3}\int_0^\infty\frac{d k}{k}
\int_0^\pi Sin^2\phi d\phi\int_{\nu_0}^\infty dy\frac{1}{k^2+4m_1^2 Cos^2\phi}
\end{equation}
\begin{displaymath}
\frac{1}{(y^2+k^2 Cos^2\phi)}\left[(1+2 Cos^2\phi)\frac{\left(1+\frac{k^2}{y^2}\right)
Cos^2\phi}{1+R(y,k^2)}+Sin^2\phi\right]F_2(y,k^2),
\end{displaymath}
where $R(y,k^2)=\sigma_L/\sigma_T$ is the ratio of the cross sections for the
transversally and longitudinally polarized virtual photons respectively. To do
numerical integration on the basis of (20) it is necessary to know the experimental
data on structure function $F_2(x,Q^2)$ and the ratio $R(x,Q^2)$ in the wide
region of variables $Q^2$ and {\it x} \cite{BK}. We have used theoretical
parameterization from \cite{AL}, where structure function $F_2$ is decomposed
into two terms $F_2^P$ and $F_2^R$, corresponding to pomeron and reggeon
exchanges:
\begin{equation}
F_2=F_2^P+F_2^R,~~~F_2^r=\frac{Q^2}{Q^2+m_0^2}C_r(t)x_r^{a_r(t)}(1-x)^{b_r(t)},
~~r=P,R,
\end{equation}
\begin{displaymath}
\frac{1}{x_r}=\frac{2m_2\nu+m_r^2}{Q^2+m_r^2}.
\end{displaymath}

\vspace{3mm}

{\bf Table. Proton polarizability contribution to the S-state \\
energy levels of electronic hydrogen and muonic hydrogen}.\\[2mm]
\begin{tabular}{|c|c|c|c|}  \hline
     &	   &	  &	 \\
     &$(e^-p^+)~~(\mu^-p^+)$	& \cite{SPH} & \cite{RR}  \\
     &Hz~~~~~~$\mu$eV  &      &      \\
     &	   &	  &	 \\    \hline
     &	   &	  &	 \\
1S   &~~~-94~~~-144~~~	   &~~~-72~~~-100~~~	  &~~~-95~~~-136~~~	 \\
     &	    &	   &	  \\	\hline
     &	    &	   &	  \\
2S   &~~~-11.8~~~-18~~~      &~~~-9~~~-13~~~	  &~~~-11.9~~~-17~~~	  \\
     &	    &	   &	   \\	 \hline
\end{tabular}

\vspace{3mm}

Recent experimental data on extraction of quantity $R(x,Q^2)$ from deep
inelastic e-p scattering made at SLAC have lead to 6-parameter models
for this ratio in a kinematic range $0.005\leq x\leq 0.86$ and
$0.5\leq Q^2\leq 130$ \cite{Abe}:
\begin{equation}
R=\frac{a_1}{\ln(Q^2/0.04)}\Theta(x,Q^2)+\frac{a_2}{[Q^8+a_3^4]^{1/4}}\left[1+
a_4 x+a_5 x^2\right] x^{a_6},
\end{equation}
\begin{displaymath}
\Theta(x,Q^2)=1+12\left(\frac{Q^2}{Q^2+1}\right)\left(\frac{0.125^2}{0.125^2+x^2}\right).
\end{displaymath}

In the case of muonic hydrogen the numerical value of Lamb shift (2P-2S)
contribution, obtained on the basis of (20)-(22), is equal to
\begin{equation}
\Delta E^{\rm Ls}_{\rm pol}=4.4\pm 0.5~~{\rm GHz},
\end{equation}
For 1S - state of electronic hydrogen the correction on proton polarizability
is $-94\pm 10 $ Hz. This contribution is in good agreement with the results
of \cite{PLH,KS,SPH,RR}.
Displacements of 1S- and 2S- energy levels of electronic hydrogen and
muonic hydrogen and the comparison with the results of \cite{SPH,RR} are
shown in Table (the displacements of muonic
hydrogen are expressed in eV for the convenience of comparison).
It is necessary to point out that the used parameterization
(21) satisfies the asymptotic condition (18) and leads to infrared finite
contribution to the energy spectrum. Thus, the main uncertainty of our result is
connected not with logarithmic approximation of the momentum integration
similar to (20) as in \cite{PLH,KS}, but with experimental errors of
$R(x,Q^2)$ in the range $Q^2\leq 0.5$ and experimental uncertainties
of $F_2(x,Q^2)$ \cite{BK}. Correction (23), obtained by us for muonic
hydrogen has the same order as other proton structure contributions
(12) and (14). Thus, it has to be taken into account when extracting of proton
charge radius $R_p$ from future experimental measurement of the muonic
hydrogen Lamb shift.

We are grateful to S.G. Karshenboim, I.B. Khriplovich, V.A. Petrun'kin,
R.A. Sen'kov, for useful discussions and to J.R. Sapirstein
for copy of paper \cite{SY}. The work was performed under the financial
support of the Russian Foundation for Fundamental Research
(grant 98-02-16185) and the Program "Universities of Russia - Fundamental
Researches" (grant 2759).

\end{document}